\documentclass[12pt,graphicx]{article}
\usepackage{amsfonts,amssymb}
\usepackage{amsthm}
\usepackage{amscd}
\input{epsf}
\date{\today}

\begin{document}

\title{\bf Tachyons on Dp-branes from Abelian Higgs sphalerons}
\author{{\large Yves Brihaye \footnote{yves.brihaye@umh.ac.be}}\\
\small{
Facult\'e des Sciences, Universit\'e de Mons-Hainaut,
B-7000 Mons, Belgium }\\
{ }\\
{\large Betti Hartmann \footnote{b.hartmann@iu-bremen.de}}\\
\small{
School of Engineering and Science, International University Bremen,
28725 Bremen, Germany
 }}

\date{\today}

\maketitle



\begin{abstract}
We consider the Abelian Higgs model in a $(p+2)$-dimensional space time
with topology $\mathbb{M}^{p+1}\times S^1$ as a field theoretical
toy model for tachyon
condensation on Dp-branes. The theory has  periodic sphaleron solutions
with the normal mode equations resembling Lam\'e-type equations.
These equations are quasi-exactly solvable (QES) for specific choices of the
Higgs- to gauge boson mass ratio and hence a finite number of algebraic
normal modes can be computed explicitely.
We calculate the tachyon potential 
for two different values of the Higgs- to gauge boson mass ratio and show that
in comparison to previously studied pure scalar field models an {\it exact}
cancellation between the negative energy contribution at the minimum 
of the tachyon potential and the brane tension is possible for the simplest
truncation in the expansion about the field around the sphaleron.
This gives further evidence for the correctness of Sen's conjecture.
\end{abstract}
\section{Introduction and Summary}
Tachyonic modes are of great interest in string theory. They have been
found to exist on D-brane--anti-D-brane pairs \cite{green} or 
non-BPS D-branes \cite{nonBPS} and are related 
to open strings ending on these D-branes.
Sen conjectured that at the minimum of the tachyon potential, the negative
contribution to the energy density from the tachyon potential should
exactly cancel the positive contribution from the tension of the D-brane--anti-D-brane
system \cite{sen}. Studies in string field theory \cite{sft} have given good hints that this
conjecture is indeed correct \cite{tachyon_sft}. In open bosonic string field
theory \cite{sz} and open superstring field theory \cite{bsz} this cancellation
works with $99\%$, respectively $85\%$ accurancy. Consequently,
toy models were searched for in which such computations would be easier than
in the string field theory models. In \cite{zwiebach} a lump solution
in a $\phi^3$-model was studied, while 
in \cite{brito} a toy model was presented which provides an instability
of Dp-branes induced by the sphaleron of a $\lambda \phi^4$ (real) scalar
field theory living in a space-time with one compactified spatial
dimension. In this model, the $Z_2$ symmetry is broken by a Higgs
potential and a non-trivial sphaleron appears at the classical level
if the field is considered to be periodic in one of the spatial
dimensions.
Technically, the treatment of the fluctuations about
the sphaleron is simplified by the fact that the fluctuation 
equation is quasi-exactly solvable (QES) \cite{turbiner}. That is to say   
that a finite number of the normal modes about the sphaleron can
be computed explicitely. It is then possible to 
check the correctness of Sen's conjecture
by restricting the phase space to
the algebraically available eigenmodes associated
with the unstable solutions.
Applying this procedure, it was shown in \cite{brito} that the cancellation between
the negative energy from the tachyon potential and the brane tension
can be nearly satisfied ($99\%$)- even when limiting to two or 
three algebraic modes only.

It is a natural question to investigate whether a similar result could
be obtained by using more elaborated theories, like gauge theories. 
Sphaleron solutions with a corresponding QES normal mode equation exist 
in the Abelian Higgs model as well. This model consists of a complex
scalar field 
coupled to electromagnetism in a $U(1)$ gauge-invariant way. The  
gauge symmetry is then broken by an appropriate Higgs potential and,
again, non-trivial sphaleron solutions occur at the classical level 
if periodicity is imposed  with respect to one of the spatial dimensions. 

In this paper, we apply the ideas and techniques of \cite{brito} to 
the Abelian Higgs model. We show 
that - even when
limiting the phase space to a single direction of instability of the
sphaleron - an {\it exact} cancellation between the negative energy contribution
from the tachyon potential and the brane tension
is possible for specific values of the radius of the compactifying circle.

In Section 2, we give the Abelian Higgs model in $(p+2)$ space-time
dimensions. In Section 3, we discuss the normal modes for 
two different values of the Higgs- to gauge boson mass ratio. In Section 4, we discuss
our results for the tachyon potential that we obtain for the two cases
considered in Section 3.

\section{The model}
\subsection{Action principle  }
We consider the Abelian Higgs model in a $(p+2)$-dimensional space-time
with topology $\mathbb{M}^{p+1}\times S^1$. The action reads:
\begin{eqnarray}
\label{lag0} S=\int   d^{p+1}y \ dx  \left[
- \frac{1}{4}  (F_{MN} F^{MN}) + (D_M \phi)^* (D^M \phi)  - V(\phi)
  \right] ,
\end{eqnarray}
where $(y^{\mu},x) = (t,y_j,x)$, $j=1,2,\dots,p$
are the world-volume's coordinates of a Dp-brane
embedded into a $p+2$ dimensional space-time and $M$, $N$ run over these coordinates. 
The coordinate $x$ is assumed to be compactified on a circle
of length $\tilde L$
such that $x\equiv x+\tilde{L}$. The complex scalar
field is denoted $\phi$ and the generalized Maxwell field $A^M$. The field
strength tensor, covariant derivative and potential read, respectively:
$$
    F_{MN}= \partial_M A_N -  \partial_N A_M \ \ \ , \ \
    D_N \phi = (\partial_N - ieA_N ) \phi \ \ , \ \
    V(\phi)= \lambda(\phi^* \phi-\frac{1}{2} v^2)^2  \ ,
$$
where $e$ is the gauge coupling constant, $v$ the vacuum expectation value
and $\lambda$ the Higgs self-coupling.

There is a spontaneous symmetry breaking of the U(1)
symmetry due to the $V(\phi)$ potential which leads to a massive gauge boson
with mass $M_W=ev$ and a massive Higgs boson $M_H=\sqrt{2\lambda} v$.

\subsection{The ansatz and classical solution}
In the following we assume the fields to be static and independent
of the brane coordinates $y_j$. We thus choose \cite{bgkk}:
\begin{equation}
\label{solution}
\phi = \frac{v}{\sqrt 2} 
\exp\left(\frac{2 i \pi x q}{\tilde L}\right) 
\Phi\left(\frac{1}{2} M_H x\right) \ , \ 
A_0 = A_{y_j}=0 \ , \ A_x = \frac{2 \pi q}{e \tilde L} \ .
\end{equation}
$\Phi(\frac{1}{2} M_H x)$ is a real function and $2q$ is an integer with $q$ 
representing the Chern-Simons charge of the solution.
In the following, we introduce the dimensionless coordinate
$z=\frac{1}{2} M_H x$ such that $L=\frac{1}{2} M_H \tilde{L}$.  
The classical equation for the field $\Phi$ reads~:
\begin{equation}
\label{kink}
              \frac{d^2}{dz^2} \Phi = 2 \Phi(\Phi^2 - 1)  \ .
\end{equation}
This equation admits periodic solutions which are determined in terms
of the Jacobi elliptic function sn$(z)$. The solution reads~:
\begin{equation}
              \Phi = k b(k) {\rm sn}(b(k) z, k) \ \ , \ \ b^2(k) =
              \frac{2}{1+k^2} \ ,
\end{equation}
where $k$  corresponds to a parameter with  $0 \leq k \leq 1$ that fixes
the period of the Jacobi function. sn($z,k$) has a period $4 K(k)$,
where $K(k)$ is the complete elliptic function of the first kind.
Specific limits are given by sn$(z,0)$ = sin$(z)$ (with $K(0)= \pi/2$),
sn$(z,1)$ = tanh$(z)$ (with $K(1)=\infty$).
As a consequence, non-trivial solutions of (\ref{kink})
exist only for $ L > L_1 \equiv \pi/\sqrt 2$.
Due to the periodicity of the function $\Phi(z)$, the function
$\phi(z)$ will be periodic with period $L$  provided
\begin{equation}
    L = \frac{2 m K(k)}{b(k)} \ \ , \ \ m \ {\rm integer}  \ 
\end{equation}
and the Chern-Simons charge $q$ of the solution should
be of the form $q= 1/2 + n$ if $m$ is odd and $q=n$ if $m$ is even.
Note that in \cite{brito} $L=\frac{4 m K(k)}{b(k)}$. We will
see in the following that the restriction to half the period
of \cite{brito} is crucial for the cancellation between the brane tension
and the negative energy from the tachyon potential.

It  turns out \cite{bgkk} that the lowest energy configuration 
(the ``sphaleron'') corresponds to $m=1$, $n=0$.
The sphaleron solution (\ref{solution}) represents the
energy barrier between vacua of different topological charges.
Its energy reads:
\begin{equation}
\label{energy}
              E_{sp} = \sqrt{ 2 \lambda } v^3 E_0(k)  \ \ {\rm with} \ \
              E_0(k) = \frac{1}{2}
               \int\limits_0^{2K(k)/b(k)} dz \left[
                     \left(\frac{d \Phi}{dz}\right)^2 + (\Phi^2 - 1)^2
                  \right]  \ .
\end{equation}
$E_0(k)$ is a monotonically increasing function of $k$:
for $k=0$ we have $E_0(0) = \pi /(4 \sqrt 2)$, while for $k=1$: $E_0(1) = 4/3$.

\section{Normal modes}
In order to determine the tachyon potential, we first have to
perform the normal mode analysis of the sphaleron solution given
above.
In the following, we use the notations of \cite{bgkk,carson}~:
\begin{equation}
\phi = \frac{v }{\sqrt 2}\exp\left(\frac{2 i \pi z q}{ L}\right)
\left(\Phi(z) + \eta_1(t,z,y_1,y_2,...,y_p) + i \eta_2(t,z,y_1,y_2,...,y_p)\right)  \ ,
\end{equation}
\begin{equation}
     A_z = \frac{2 \pi q}{e  L} + \frac{v}{e} a_z(t,z,y_1,y_2,...,y_p) \ \ , \ \
     A_{\mu} = \frac{v}{e} a_{\mu}(t,z,y_1,y_2,...,y_p) \ \ , \  \mu=0, y_1, y_2,..., y_p \ ,
\end{equation}
where the functions $a_z$, $a_{\mu}$ are periodic with respect to $z$ on $[0,L]$,
while the functions $\eta_1$ and $\eta_2$ are periodic, 
respectively anti-periodic with respect to $z$
if $2q$ is even and $2q$ is odd. 

We fix the gauge degree of freedom by choosing the 
background gauge condition:
\begin{equation}
       G(a,\eta) = 
       \frac{i}{2} \theta(\eta^* \Phi - \eta \Phi)-\partial_{M} a^{M} 
                 = 0 \ \ {\rm with} \ \ M=t,z,y_1,..,y_p \ ,
\end{equation}
where $\theta \equiv 2M_H/M_W = \sqrt{2 \lambda}/e$ and $\eta\equiv (\eta_1,\eta_2)$.

Expanding the classical action in powers of the fluctuations $a_z$, $a_{\mu}$,
$\eta_1$ and $\eta_2$ 
and using the gauge fixing condition leads to the following
expression for the action~:

\begin{eqnarray}
\label{expansion}
  S &=& \frac{1}{2} \int d^{p+1}y \ dz \left[
  -\left(\frac{d \Phi}{dz}\right)^2 - (\Phi^2 - 1)^2
   - \partial_{M}\eta   \partial^{M}\eta
   -  \eta^{\dagger} H(\Phi) \eta  - \Lambda_{inter}
   \right]
\end{eqnarray}
where  the quadractic term $\eta^{\dagger} H(\Phi) \eta$ has the form~:
 \begin{eqnarray}
   \eta^{\dagger} H(\Phi) \eta = \eta_1 \left[\partial_z^2 - (6\Phi^2 -2)\right]\eta_1
   + \sum_{\mu} a_{\mu}( \partial_z^2 - \theta ^2\Phi^2) a_{\mu}
   + \left(\matrix{a_z \ \eta_2}\right )  M_2 \left(\matrix{a_z \cr \eta_2\cr
              }\right)
 \end{eqnarray}
 and $M_2$ is the  operator  matrix defined in (\ref{system}) below.
The ``interaction term'' $\Lambda_{inter}$ reads~:
 \begin{eqnarray}
 \Lambda_{inter} &=&  4 \Phi \eta_1 ( \eta_1^2 + \eta_2^2) + (\eta_1^2 + \eta_2^2)^2    
   + a_z^2 (\eta_1^2 + \eta_2^2)  +  2 a_z^2 \Phi \eta_1  \nonumber \\ 
   &+& a_z \frac{d}{dz} (\eta_1^2 + \eta_2^2)
+ a_{\mu} \partial^{\mu} ( \eta_1^2 + \eta_2^2) 
+ 2 \eta_1 \Phi a_{\mu}a^{\mu}  \ .
\end{eqnarray}

The $dz$-integral of the first two terms in (\ref{expansion}) is the negative of the 
(rescaled) energy $E_0$ 
of the sphaleron solution (compare (\ref{energy})) and is equal to the Dp-brane 
tension $T_p\equiv -E_0$.

We use the following normal mode expansion:
\begin{eqnarray}
\eta_s(t,z,y_1,..,y_p)=\sum\limits_{n_{(s)}} \xi^{(s)}_{n_{(s)}}(y) \psi^{(s)}_{n_{(s)}}(z) \ , \ \ s=1,2 
\end{eqnarray}
and
\begin{eqnarray} 
a_M(t,z,y_1,..,y_p)
=\sum\limits_{n_{(\small{M})}} \zeta^{(\small{M})}_{n_{(\small{M})}}(y) \chi^{(M)}_{n_{(M)}}(z) \ , \ M=0,z,y_1,..,y_p \ . 
\end{eqnarray}
The fields $\xi^{(s)}_{n_{(s)}}(y)$  and $\zeta^{(\small{M})}_{n_{(\small{M})}}(y)$ live on the
Dp-brane world volume and satisfy the equations
\begin{equation}
\square_{p+1}\xi^{(s)}_{n_{(s)}}(y)=\omega^2 \xi^{(s)}_{n_{(s)}}(y) \ \ , \ \
\square_{p+1}\zeta^{(M)}_{n_{(M)}}(y)=\omega^2 \zeta^{(M)}_{n_{(M)}}(y) \ ,
\end{equation}
while the expressions for the fields $\psi^{(s)}_{n_{(s)}}(z)$ and $\chi^{(M)}_{n_{(M)}}(z)$  lead to the following system
of Schr\"odinger-like equations~:
\begin{equation}
\label{lame1}
   (-\partial_z^2 + 6 \Phi^2 - 2) \psi^{(1)}_{n_{(1)}} = \omega^2_{n_{(1)}}  \psi^{(1)}_{n_{(1)}}  \ ,
\end{equation}
\begin{equation}
\label{lame2}
   (-\partial_z^2 + \theta^2 \Phi^2) \chi^{(\mu)}_{n_{(\mu)}}  = 
\omega^2_{n_{(\mu)}} \chi^{(\mu)}_{n_{(\mu)}}  \ ,  \  \mu=t,y_1,y_2,...,y_p 
\end{equation}
and
\begin{equation}
\label{system}
\left(\matrix{-\partial_z^2 + \theta^2 \Phi^2 & 2\theta \Phi' \cr
              2\theta \Phi' &   -\partial_z^2 + (\theta^2+2) \Phi^2 -2
              }\right)
\left(\matrix{\chi^{(z)}_{n_{(z)}}    \cr \psi^{(2)}_{n_{(2)}}    \cr
              }\right)
= \omega^2_{n_{(z)}} 
\left(\matrix{\chi^{(z)}_{n_{(z)}} \cr \psi^{(2)}_{n_{(2)}}\cr
              }\right)   \ .
\end{equation}
Equation (\ref{lame1}) above is a Lam\'e equation
and admits five explicit eigenvalues. The equations (\ref{lame2})
and (\ref{system}) do not admit explicit solutions for generic
values of mass ratio  $\theta$. However if this mass ratio is of the
form $\theta^2 = N(N+1)$ (with $N$ being an integer) then (\ref{lame2})
is a  Lam\'e equation admitting  $2N+1$ explicit solutions (for each value of
the index $\mu$)  and the coupled  equation (\ref{system}) admits   
$4N+2$ explicit solutions (for $N\ge 1$) \cite{bgkk,bb}. 
 
In the following, we will discuss the cases $N=1$ and $N=2$, i.e. $\theta^2 =2$
and $\theta^2=6$, respectively, in detail.
\subsection{$N=1$}
For $N=1$ we have a total of $14+3p$ explicit eigenvalues of the quadratic 
form about the sphaleron.
Note that, of course, there are more (non-explicit) eigenvectors
but along with \cite{brito} we will only discuss the algebraically available here.
In the following, we will give the possible algebraically available solutions. 
 \begin{enumerate}
  \item $\psi^{(1)}$-channel \cite{brito}:
\begin{eqnarray*}
\psi^{(1)}_{0} &=& {\rm sn}^2 - 
\frac{1}{3k^2}(1+k^2 + \sqrt{1-k^2+k^4}) \ \ , \ \  \omega^2_0=\left(1+k^2-2\sqrt{1-k^2+k^4}\right)b(k)^2 \nonumber\\
\psi^{(1)}_{1} &=& {\rm cn}\ {\rm dn} \ \ ,  \ \  \omega^2_1=0 \nonumber \\
\psi^{(1)}_{2} &=& {\rm sn}\ {\rm dn} \ \ , \ \  \omega^2_2=3 k^2 b(k)^2 \nonumber \\ 
\psi^{(1)}_{3} &=& {\rm sn}\ {\rm cn} \ \ , \ \  \omega^2_3=3 b(k)^2 \nonumber \\
\psi^{(1)}_{4} &=& {\rm sn}^2-\frac{1}{3 k^2}\left(1+k^2-\sqrt{1-k^2+k^4}\right) \ \ , \ \
\omega^2_4=\left(1+k^2+2\sqrt{1-k^2+k^4}\right)b(k^2) 
\end{eqnarray*}
  \item $\psi^{(2)}$-channel \cite{bgkk}:
\begin{eqnarray*}
\psi^{(2)}_{0} &=& \sqrt{2}(\Phi^2-1)  \ \ , \ \  \omega^2_0=-2 \nonumber\\
\psi^{(2)}_{1} &=& \sqrt{2}\Phi' \ \ ,  \ \  \omega^2_1=-2kb(k)^2 \nonumber \\
\psi^{(2)}_{2} &=& -\frac{\sqrt{2}}{k}\Phi  \ {\rm dn} \ \ , \ \  \omega^2_2= k^2 b(k)^2  \nonumber \\ 
\psi^{(2)}_{3} &=& \sqrt{2}\Phi' \ \ , \ \  \omega^2_3= 2kb(k)^2\nonumber \\
\psi^{(2)}_{4} &=& -\sqrt{2}k \ \Phi \ {\rm cn} \ \ , \ \ \omega^2_4= b(k)^2 \nonumber \\
\psi^{(2)}_{5} &=& \sqrt{2} \Phi^2 \ \ , \ \
\omega^2_5= 2
\end{eqnarray*}
\item  $\chi^{(\mu)}_{n_{(\mu)}}$-channel:
\begin{eqnarray*}
\chi^{(\mu)}_{0} &=&  {\rm sn} \ \  , \ \ \omega^2_0 = 1+k^2 \nonumber \\
\chi^{(\mu)}_{1} &=&  {\rm cn} \ \  , \ \ \omega^2_1 = 1 \nonumber \\
\chi^{(\mu)}_{2} &=&  {\rm dn} \ \  , \ \ \omega^2_2 = k^2 
\end{eqnarray*}
\item $\chi^{(z)}_{n_{(z)}}$-channel \cite{bgkk} :
\begin{eqnarray*}
\chi^{(z)}_{0} &=&  \Phi' \ \  , \ \ \omega^2_0 =-2 \nonumber \\
\chi^{(z)}_{1} &=&  \Phi^2+\frac{1}{2}\omega^2_1 \ \  , \ \ \omega^2_1 = -2kb(k)^2 \nonumber \\
\chi^{(z)}_{2} &=&  \Phi \ {\rm cn} \ \  , \ \ \omega^2_2 = k^2 b(k)^2 \nonumber \\
\chi^{(z)}_{3} &=&  \Phi^2 +\frac{1}{2}\omega^2_3 \ \  , \ \ \omega^2_3 = 2kb(k)^2 \nonumber \\
\chi^{(z)}_{4} &=&  \Phi \ {\rm dn} \ \  , \ \ \omega^2_4 = b(k)^2 \nonumber \\
\chi^{(z)}_{5} &=&  \Phi' \ \  , \ \ \omega^2_5 = 2 \ ,
\end{eqnarray*}
\end{enumerate}
where we have used ${\rm sn}$, ${\rm cn}$ and ${\rm dn}$ as abbreviation
for the Jacobi elliptic functions ${\rm  sn}(b(k)z,k)$, 
${\rm  cn}(b(k)z,k)$ and ${\rm  dn}(b(k)z,k)$, respectively. 
The above given eigenvectors are not normalised. We will introduce
a normalisation in the computations below.
   
Note that not all given normal modes are of interest for our study:
since we study the sphaleron solution with $q=1/2$ ($2q$ odd) here, 
the functions $\eta_1$ and $\eta_2$ have to be anti-periodic on $[0,L]$, while
there is no restriction on the $a_M$ fields. Note, however, that the
$\psi^{(2)}$ channel and the $\chi^{(z)}$ channel are directly linked.
Moreover, from
(\ref{solution}) we find that the function $\Phi(z)$ has
also to be anti-periodic on the interval $[0,L]$. Not
all normal modes given above possess this property. 
Let us discuss this in more detail:
\begin{itemize}
\item $\psi^{(1)}$-channel:
Only the modes $\psi^{(1)}_1$ and $\psi^{(1)}_2$  are anti-periodic. 
$\psi^{(1)}_{1}$ is a zero-mode and $\psi^{(1)}_{2}$
a positive mode. So, no negative mode which is anti-periodic appears 
in this channel. This contrasts
with the situation in \cite{brito} where the  solution
 $\psi^{(1)}_{0}$ plays a central role since it is the negative mode
and periodic as required in \cite{brito}.

\item  $\psi^{(2)}$- and $\chi^{(z)}_{n_{(z)}}$-channel:
Only the modes $\psi^{(2)}_{1}$, $\psi^{(2)}_{2}$, 
$\psi^{(2)}_{3}$ (and related to that 
$\chi^{(z)}_{1}$, $\chi^{(z)}_{2}$, $\chi^{(z)}_{3}$)
are of interest for us since they are anti-periodic. However,
only $\psi^{(2)}_{1}$ is a negative mode, i.e. a tachyonic mode, and thus the (only) one of interest for us.
\end{itemize}

\subsection{$N=2$}
For $N=2$ we have a total of $20+5p$ explicit eigenvalues:
\begin{enumerate}
\item $\psi^{(1)}$-channel: Since (\ref{lame1}) is independent on $N$ the eigenvalues and functions
are the same for all $N$. 
\item $\psi^{(2)}$-channel \cite{bgkk}:
\begin{eqnarray*}
\psi^{(2)}_{0} &=& \sqrt{6}\left[2\Phi' {\rm cn}-\frac{\Phi^2 \ {\rm dn}}{k}
- \ {\rm dn}\left(\frac{\omega_0^2-3b(k)^2}{6k}\right) \right] \ \ , \ \  \omega^2_0=b(k)^2(1-2\sqrt{1+3k^2}) \nonumber\\
\psi^{(2)}_{1} &=&  \sqrt{6}\left[2\Phi' {\rm dn}-\Phi^2 \ {\rm dn} \ k
- \frac{k}{6} \  {\rm cn}\left(\omega^2_1-3b(k)^2 k^2\right)\right]
\  ,   \  \omega^2_1=k^2 b(k)^2 \left(1-\frac{2}{k}\sqrt{k^2+3}\right) \nonumber \\
\psi^{(2)}_{2} &=& \sqrt{6}\left(\Phi^3 -(8-\omega^2_2)\Phi/6\right) \ \ , \ \  \omega^2_2=4-2b(k)^2\sqrt{1-k^2+k^4}  \nonumber \\
\psi^{(2)}_{3} &=&  \sqrt{6}\left[2\Phi' \ {\rm dn}-\Phi^2 \ {\rm dn} \ k
- \frac{k}{6} \ {\rm cn}\left(\omega^2_3-3b(k)^2 k^2\right)\right] \  , \  \omega^2_3= k^2 b(k)^2 
\left(1+\frac{2}{k}\sqrt{k^2+3}\right)\nonumber \\
\psi^{(2)}_{4} &=& \frac{\sqrt{6}}{2}\Phi\Phi' \ \ ,  \ \  \omega^2_4=2 \nonumber \\
\psi^{(2)}_{5} &=& -\sqrt{6} \Phi^2 \ {\rm dn} \ \ , \ \ \omega^2_5= b(k)^2(1+4k^2) \nonumber \\
\psi^{(2)}_{6} &=&  \sqrt{6}\left[\left(2\Phi' \ {\rm cn}
-\frac{\Phi^2 \ {\rm dn}}{k}\right)-{\rm dn}\left(\frac{\omega^2_6-3b(k)^2}{6k}\right) \right]
\ \ , \ \ \omega^2_6= b(k)^2(1+2\sqrt{1+3 k^2})  \nonumber \\
\psi^{(2)}_{7} &=& \frac{\sqrt{6}}{2}\Phi\Phi' \ \ ,  \ \  \omega^2_7=6 \nonumber \\
\psi^{(2)}_{8} &=& \sqrt{6}\left[\Phi^3 -(8-\omega^2_8)\Phi/6\right]   \ \ , \ \  \omega^2_8=4+ 2b(k)^2\sqrt{1-k^2+k^4}  \nonumber \\
\psi^{(2)}_{9} &=& -\sqrt{6} k \ \Phi^2 \ {\rm cn} \ \ , \ \  \omega^2_9= b(k)^2(4+k^2)
\end{eqnarray*}
\item  $\chi^{(\mu)}_{n_{(\mu)}}$-channel:
\begin{eqnarray*}
\chi^{(\mu)}_{0} &=& {\rm sn}^2 -
\frac{1}{3k^2}(1+k^2 + \sqrt{1-k^2+k^4}) \ \ , \ \  \omega^2_0=\left(1+k^2-2\sqrt{1-k^2+k^4}\right)b(k)^2+2 \nonumber\\
\chi^{(\mu)}_{1} &=& {\rm cn}\ {\rm dn} \ \ ,  \ \  \omega^2_1=2 \nonumber \\
\chi^{(\mu)}_{2} &=& {\rm sn}\ {\rm dn} \ \ , \ \  \omega^2_2=3 k^2 b(k)^2 +2 \nonumber \\ 
\chi^{(\mu)}_{3} &=& {\rm sn}\ {\rm cn} \ \ , \ \  \omega^2_3=3 b(k)^2 +2 \nonumber \\
\chi^{(\mu)}_{4} &=& {\rm sn}^2-\frac{1}{3 k^2}\left(1+k^2-\sqrt{1-k^2+k^4}\right) \ \ , \ \
\omega^2_4=\left(1+k^2+2\sqrt{1-k^2+k^4}\right)b(k^2) +2
\end{eqnarray*}
\item $\chi^{(z)}_{n_{(z)}}$-channel \cite{bgkk} :
\begin{eqnarray*}
\chi^{(z)}_{0} &=&  6\Phi^2 \ {\rm cn} + (\omega^2_0-3b(k)^2)\ {\rm cn} \ \  , \ \ \omega^2_0 =b(k)^2(1-2\sqrt{1+3k^2}) \nonumber \\
\chi^{(z)}_{1} &=&  6\Phi^2 \ {\rm dn} + (\omega^2_1-3b(k)^2 k^2) \ {\rm dn}  \ \  , \ \ \omega^2_1 =k^2 b(k)^2 \left(1-\frac{2}{k}\sqrt{k^2+3}\right)  \nonumber \\
\chi^{(z)}_{2} &=&  2\Phi\Phi' \ \  , \ \ \omega^2_2 = 4-2b(k)^2\sqrt{1-k^2+k^4}  \nonumber \\
\chi^{(z)}_{3} &=&   6\Phi^2 \ {\rm dn} + (\omega^2_3-3b(k)^2 k^2) \ {\rm dn} \ \  , \ \ \omega^2_3 = k^2 b(k)^2\left(1+ \frac{2}{k}\sqrt{k^2+3}\right)  \nonumber \\
\chi^{(z)}_{4} &=&  \Phi^3-\Phi \ \  , \ \ \omega^2_4 = 2 \nonumber \\
\chi^{(z)}_{5} &=&  2k\Phi^2 \ {\rm cn} - k b(k)^2 \ {\rm cn} \ \  , \ \ \omega^2_5 =   b(k)^2(1+4k^2)\nonumber \\
\chi^{(z)}_{6} &=&  6\Phi^2 \ {\rm cn} + (\omega^2_6-3b(k)^2) \ {\rm cn} \ \  , \ \ \omega^2_6 =   b(k)^2(1+2\sqrt{1+3 k^2}) \nonumber \\
\chi^{(z)}_{7} &=&  \Phi^3 \ \  , \ \ \omega^2_7 =   6 \nonumber \\
\chi^{(z)}_{8} &=&   2\Phi\Phi' \ \  , \ \ \omega^2_8 =   4+ 2b(k)^2\sqrt{1-k^2+k^4} \nonumber \\
\chi^{(z)}_{9} &=&  2\Phi^2 \ {\rm dn} - k^2 b(k)^2 \ {\rm dn} \ \  , \ \ \omega^2_9 =   b(k)^2(4+k^2) \nonumber \\
\end{eqnarray*}

\end{enumerate}
Again, only the $\psi^{(2)}_1$ mode (and with that $\chi_1^{(z)}$)
is of interest for us since it is both tachyonic and anti-periodic. 

\section{Tachyon potential}
 Following the ideas of \cite{brito}, we will now discuss the computation of the
tachyon potential. For $N=1$ we write the discrete modes as:
\begin{eqnarray}
\eta_1&=&\xi^{(1)}_0 \psi^{(1)}_0+\xi^{(1)}_1 \psi^{(1)}_1+ \xi^{(1)}_2 \psi^{(1)}_2+
 \xi^{(1)}_3 \psi^{(1)}_3+\xi^{(1)}_4 \psi^{(1)}_4 \ , \nonumber \\
\eta_2&=&\xi^{(2)}_0 \psi^{(2)}_0+\xi^{(2)}_1 \psi^{(2)}_1+ \xi^{(2)}_2 \psi^{(2)}_2+
 \xi^{(2)}_3 \psi^{(2)}_3+\xi^{(2)}_4 \psi^{(2)}_4 +\xi^{(2)}_5 \psi^{(2)}_5  \ , \nonumber \\
a_{\mu}&=&\zeta^{(\mu)}_0 \chi^{(\mu)}_0
+\zeta^{(\mu)}_1 \chi^{(\mu)}_1+\zeta^{(\mu)}_2 \chi^{(\mu)}_2 \ , \nonumber \\
a_{z}&=&\xi^{(2)}_0 \chi^{(z)}_0
+\xi^{(2)}_1 \chi^{(z)}_1+\xi^{(2)}_2 \chi^{(z)}_2 
+\xi^{(2)}_3 \chi^{(z)}_3+\xi^{(2)}_4 \chi^{(z)}_4 +\xi^{(2)}_5 \chi^{(z)}_5  \ ,
\end{eqnarray}
where we have identified $\xi^{(2)}_i\equiv \zeta^{(z)}_i$, $i=0,..,5$, since the two channels are linked.

For $N=2$, we have
\begin{eqnarray}
\eta_1&=&\xi^{(1)}_0 \psi^{(1)}_0+\xi^{(1)}_1 \psi^{(1)}_1+ \xi^{(1)}_2 \psi^{(1)}_2+
 \xi^{(1)}_3 \psi^{(1)}_3+\xi^{(1)}_4 \psi^{(1)}_4 \ , \nonumber \\
\eta_2&=&\xi^{(2)}_0 \psi^{(2)}_0+\xi^{(2)}_1 \psi^{(2)}_1+ \xi^{(2)}_2 \psi^{(2)}_2+
 \xi^{(2)}_3 \psi^{(2)}_3+\xi^{(2)}_4 \psi^{(2)}_4 +\xi^{(2)}_5 \psi^{(2)}_5 +
 \xi^{(2)}_6 \psi^{(2)}_6+\xi^{(2)}_7 \psi^{(2)}_7 \nonumber \\
&+&\xi^{(2)}_8 \psi^{(2)}_8
+\xi^{(2)}_9 \psi^{(2)}_9
\ , \nonumber \\
a_{\mu}&=&\zeta^{(\mu)}_0 \chi^{(\mu)}_0
+\zeta^{(\mu)}_1 \chi^{(\mu)}_1+\zeta^{(\mu)}_2 \chi^{(\mu)}_2 
+\zeta^{(\mu)}_3 \chi^{(\mu)}_3+\zeta^{(\mu)}_4 \chi^{(\mu)}_4 
\ , \nonumber \\
a_{z}&=&\xi^{(2)}_0 \chi^{(z)}_0
+\xi^{(2)}_1 \chi^{(z)}_1+\xi^{(2)}_2 \chi^{(z)}_2 
+\xi^{(2)}_3 \chi^{(z)}_3+\xi^{(2)}_4 \chi^{(z)}_4 +\xi^{(2)}_5 \chi^{(z)}_5  
+\xi^{(2)}_6 \chi^{(z)}_6 
+\xi^{(2)}_7 \chi^{(z)}_7 \nonumber \\
&+&\xi^{(2)}_8 \chi^{(z)}_8 +\xi^{(2)}_9 \chi^{(z)}_9 \ ,
\end{eqnarray}
where -similar to $N=1$- we have identified $\xi^{(2)}_i\equiv \zeta^{(z)}_i$, $i=0,..,9$.

In the following $\eta_1$ will not contribute since none of the normal modes is both anti-periodic
and negative. For $a_{\mu}$ all modes are positive and are thus also not of interest for us. 
For $\eta_2$ (respectively $a_z$) only the $\xi^{(2)}_1$ mode will contribute.
After substituting the normal mode expansion into (\ref{expansion}) and integrating out
all modes over a period $2K(k)/b(k)$ we thus obtain the following
 effective action for the Dp-brane:
  \begin{equation}
                S_p = \int d^{p+1}y \left [
           -T_p - \frac{1}{2}
           \partial_{\mu}\xi^{(2)}_{1}(y)  \partial^{\mu}\xi^{(2)}_{1}(y)
           -V(\xi)  \right ]
  \end{equation}
  with $T_p$ representing the tension of the Dp-brane and the
  effective potential: 
  \begin{equation}
\label{potential}
  V(\xi) = \frac{1}{2} \left[-\omega^2_1 \xi^{(2)}_1
  - B(k) (\xi^{(2)}_1)^4 \right]  \ ,   \  B(k) = 
\int\limits_0^{2 K(k)/b(k)} \nu^4 \left[4 (\Phi')^4 +2 (\Phi')^2 
\left(\Phi^2+\frac{\omega_1^2}{2}\right)  \right] dz \ ,
  \end{equation}
where the prime denotes the derivative with respect to $z$.
$\omega^2_1 = -2 k b(k)^2$ is the negative mode of the
  sphaleron, while $B(k)$ depends on $k$. $\nu$ is the normalization of the corresponding eigenvector.

We have used here the normalisation and orthogonality of the eigenfunctions:
\begin{equation}
\int\limits_{0}^{2K(k)/b(k)} dz \ \psi_n(z) \psi_m(z) = \delta_{n,m} \ , \ 
\int\limits_{0}^{2K(k)/b(k)} dz \ \chi_n(z) \chi_m(z) = \delta_{n,m} \ .
\end{equation}
For tachyon condensation we require - following Sen's conjecture - that
the contribution from the tachyon potential exactly cancels the brane tension  at the critical point $\xi^*$ :
\begin{equation}
T_p+V(\xi^*)=0 \ .
\end{equation}

\begin{figure}
\epsfysize=18cm
\centering
\epsffile{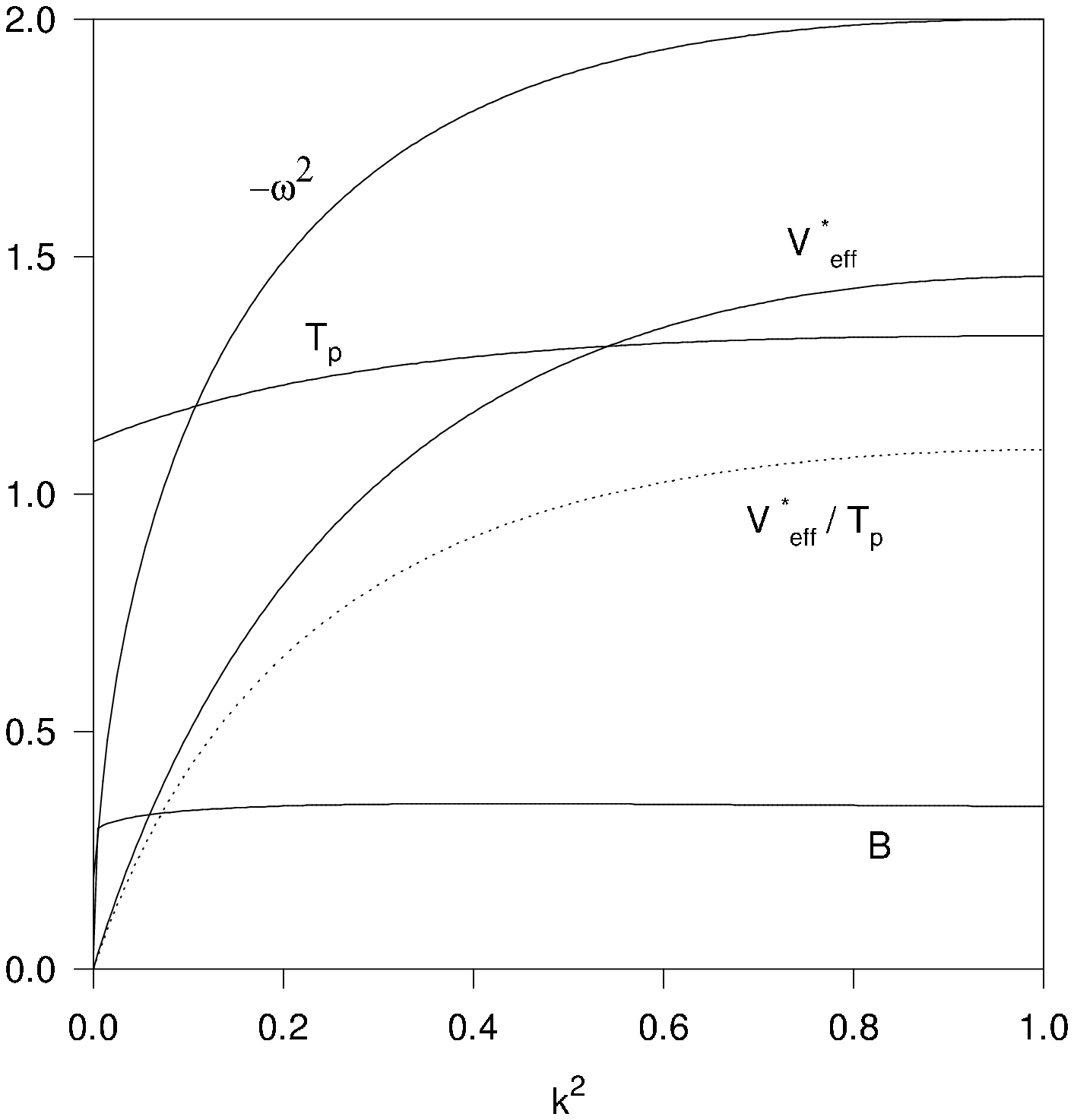}
\caption{\label{fig1} 
The brane tension $T_p$, the modulus of the potential value of the critical value $V_{eff}^*\equiv \vert
V(\xi^*)\vert$,
the negative of the tachyonic eigenmode of the sphaleron $\omega^2\equiv \omega^2_1$ and the integral $B(k)$
 are given as  functions of
$k^2$ for gauge- to Higgs boson mass $=\sqrt{2}$ ($N=1$). The ratio
$V_{eff}^*/T_p$ is also shown.
}
\end{figure}

\begin{figure}
\epsfysize=18cm
\centering
\epsffile{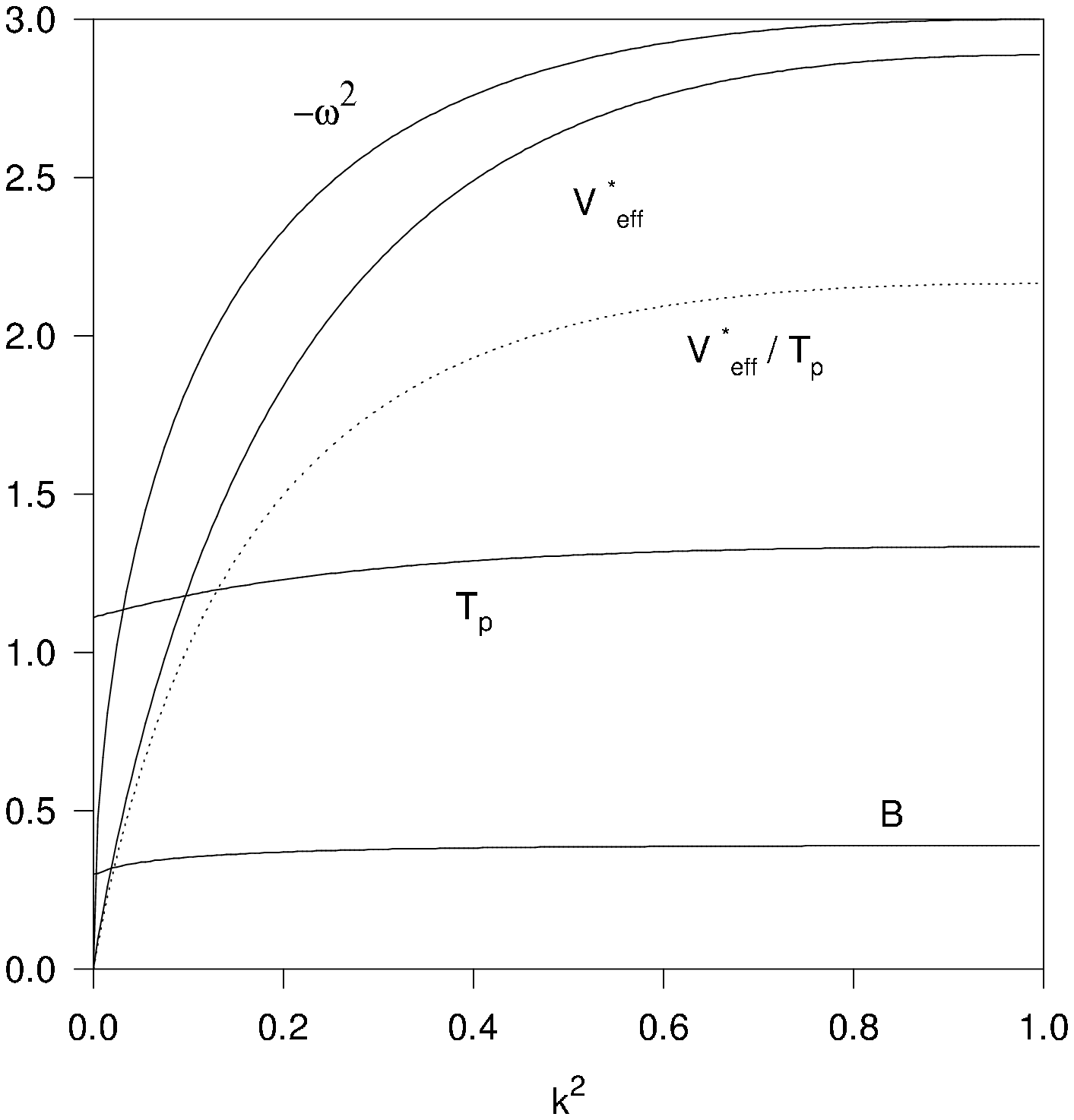}
\caption{\label{fig2} 
Same as Fig.\ref{fig1} for gauge- to Higgs boson mass $=\sqrt{2/3}$ ($N=2$).}
\end{figure}

\begin{figure}
\epsfysize=18cm
\centering
\epsffile{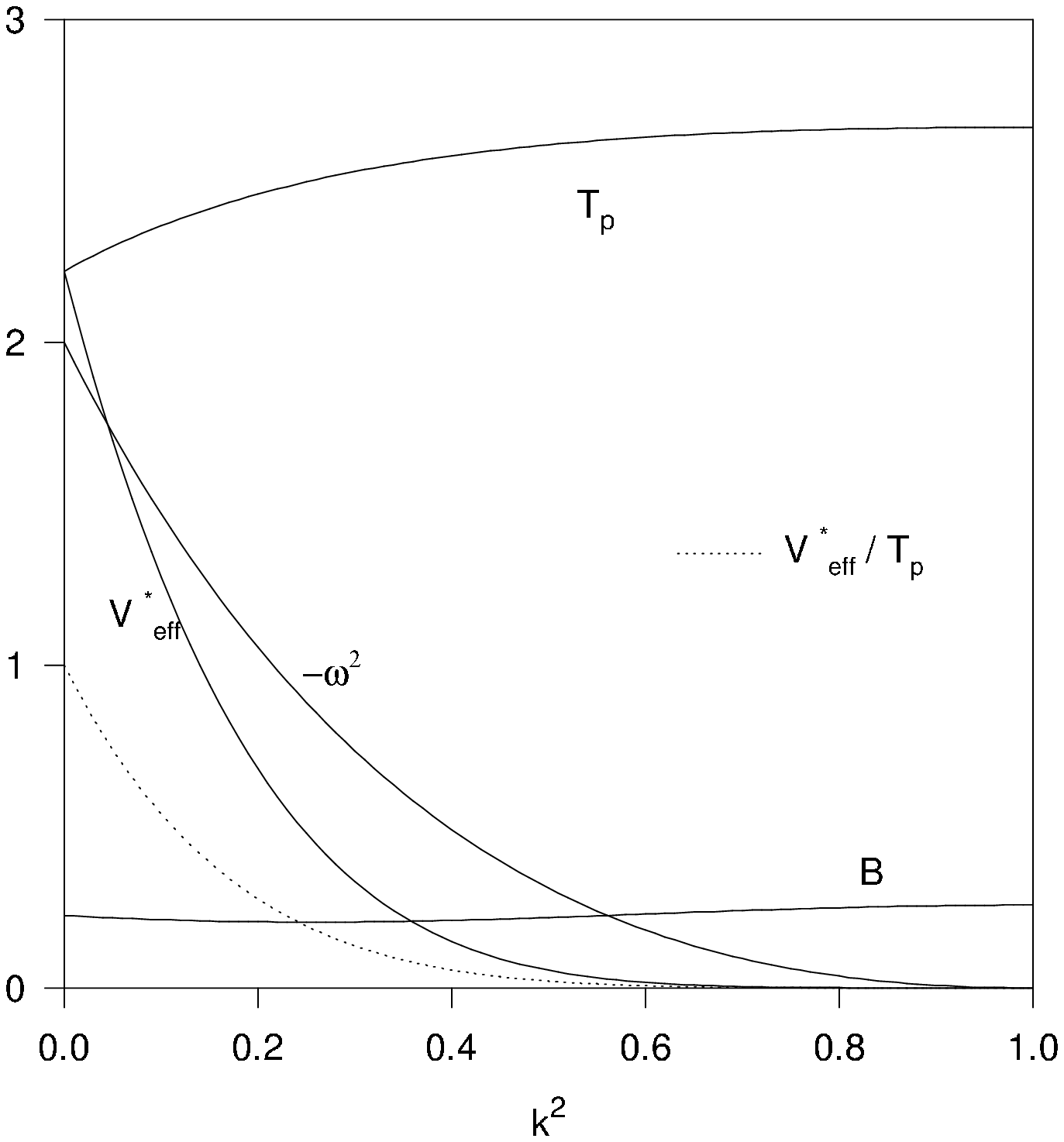}
\caption{\label{fig3} 
Same as Fig.\ref{fig1} for the model involving the $\phi^4$
scalar model of \cite{brito}.}
\end{figure}

Evaluation of  the
  critical value $(\xi^{(2)}_1)^*\equiv \xi^*$ of (\ref{potential}) is then straightforward.
  The string tension $T_p$, the modulus of the minimal value of the 
potential $V_{eff}^*\equiv \vert
V(\xi^*)\vert$ 
  and the values $-\omega^2\equiv -\omega^2_1$, $B(k)$ are shown 
  as functions of $k^2$ for $N=1$ in Fig.~\ref{fig1} and for $N=2$ in Fig.~\ref{fig2}. 
Also shown is the ratio
  $V_{eff}^*/T_p=\vert V(\xi^*)\vert/T_p $. Remarkably, our results indicate that for 
  $k^2 \approx 0.64$ ($N=1$), respectively $k^2\approx 0.097$ ($N=2$) 
the brane tension is equivalent to the value of the tachyon
potential at the minimum, i.e.  $V_{eff}^*/T_p =\vert V(\xi^*)\vert/T_p  = 1$. 
For comparison, we also present the corresponding results for the case of a ``pure'' scalar
 studied in \cite{brito}. The data is shown in Fig.\ref{fig3}.
The figure demonstrates that in the scalar sphaleron case using the
mode approximation   $V_{eff}^*/T_p=\vert V(\xi^*)\vert/T_p  = 1$ is only satisfied in the limit
$k^2 = 0$ (in which case the scalar sphaleron becomes a trivial function
since $\Phi(z) = 0$) and our numerical results agree with those in \cite{brito}.
We notice that the crucial difference between our results and those of \cite{brito}
is that $\omega(k)$ decreases (as function
of $k$) in our case, while it increases in \cite{brito}.
It is this difference which allows for the effective potential
to become equal to the string tension at some non-trivial value of $k$.
\cite{brito}  suggests his results could be improved by inclusion of
two or more further eigenmodes. We could perform a similar analysis here, however,
since the exact cancellation works for a finite value of $k^2$, the correctness
of Sen's conjecture is shown already by studying only one normal mode about the sphaleron.\\
\\
{\bf Acknowledgements}\\
YB is grateful to the
Belgian FNRS for financial support.
%


\newpage

\end{document}